\documentstyle[twoside,fleqn,espcrc2]{article}

\title{\Large \bf Phenomenological constraints\\
on a scale-dependent gravitational coupling\thanks{Talk presented by one of us
(O.B.) at the Workshop on Theoretical and Phenomenological Aspects of
Underground Physics (TAUP 95), Toledo, Spain, September 1995; to appear
in Nucl. Phys. B. supplement.}}

\author{Orfeu Bertolami\address{\it Departamento de F\'\i sica,
          Instituto Superior T\'ecnico, Av. Rovisco Pais,
          1096 Lisboa Codex, Portugal} %
         and
        Juan Garc\'\i a--Bellido\address{\it Astronomy Centre,
        University of Sussex, Brighton BN1 9QH, UK}}

\begin{document}

\begin{abstract}
We investigate the astrophysical and cosmological implications
of the recently proposed idea
of a running gravitational coupling on macroscopic scales. We find that
when applied
to the rotation curves of galaxies,  their flatness requires still
the presence of dark matter. Bounds on the variation of the
gravitational coupling from primordial nucleosynthesis, change of
the period of binary pulsars, gravitational lensing and the
cosmic virial theorem are analysed.
\end{abstract}

\maketitle

\section{INTRODUCTION}

The flatness of the rotation curves of galaxies and the large structure
of the Universe indicate that either the Universe is predominantly made
up of dark matter of exotic nature, i.e. non-baryonic, and/or that on
large scales gravity is distinctively different from that on solar
system scales, where Newtonian and post-Newtonian approximations are
valid. The former possibility has been thoroughly investigated
(see Ref. \cite{Trimble}
for a review) and is an active subject of research in
astroparticle physics. The second possibility, however relevant, has drawn
less attention. This
is essentially due to the fact that until recently no consistent and
appealing modification of Newtonian and post-Newtonian dynamics has been put
foward. Many of these attempts \cite{Finzi},
although consistent with observations, were most
often unsatisfactory from the theoretical point of view. Actually, it has
been recently shown that under certain fairly general
conditions it is unlikely that
relativistic gravity theories can explain the flatness of the rotation curves
of galaxies \cite{Nester}.
These conditions however do not exclude the class of
generalizations of General Relativity that involve higher-derivatives.
Quantum versions of these theories were shown to exhibit asymptotic
freedom in the gravitational coupling \cite{Julve} and one would expect
this property to manifest itself on large scales. This possibility
would surprisingly imply that quantum effects could mimic the
presence of dark matter \cite{Goldman}, as well as induce other
cosmological phenomena \cite{Bertolami,Bertolami1}. One
striking implication of these ideas is the prediction
\cite{Bertolami,Bertolami1} that the
power spectrum on large scales would have more power than the one predicted
by the $\Omega = 1$ Cold Dark Matter (CDM) Model, in agreement with what is
observed by IRAS \cite{IRAS}. Furthermore, due to the increase in the
gravitational constant on large scales one finds that
the energy density fluctuations grow quicker than in usual
matter dominated Friedmann-Robertson-Walker models
\cite{Bertolami,Bertolami1}. Moreover, one
can explain with a scale-dependent $G$ the discrepancy between
determinations of the Hubble's parameter made at different scales,
as suggested in \cite{Bertolami}, and studied in \cite{Kim}.

Nevertheless, independently of the possible running of the gravitational
constant in a higher derivative theory of gravity, it is
worthwhile analysing the constraints on the scale-dependence of $G$ from
astrophysical and cosmological phenomena, where such an effect would
be dominant. On the other hand, in the last few years there has been a
revival of Brans-Dicke like theories, with variable gravitational coupling,
that has led to a number of constraints on possible time
variations of $G$. Of course, some of the constraints on $\dot G$ can be
written as constraints on $\Delta G$ over scales in which a graviton
took a time $\Delta t$ to propagate. For instance, during
nucleosynthesis the largest distance that a graviton could have
traversed is the horizon distance at that time, {\it i.e.} a
few ligh-seconds to a few light-minutes,
approximately the Earth-Moon distance. Such a distance is too small
for quantum effects to become appreciable, as we  discuss below.
However, those effects become important at kiloparsec (kpc) distances
and therefore could be relevant for discussing the rotation curves of
galaxies. We shall actually show, for a particular theory
\cite{Goldman,Bertolami}, that
the rotation curves of spiral galaxies cannot be entirely explained by
the running of $G$, so some amount of dark matter is required,
which is still consistent with the upper bound on baryonic matter coming
from primordial nucleosynthesis.
On the other hand, we could impose bounds on a possible variation of $G$
from a plethora of cosmological and astrophysical phenomena at large
scales, although the lack of precise observations at those scales
make the bounds rather weak \cite{BGB}. Of course, a difficulty
in examining constraints on the variation of $G$ is that
in all gravitational phenomena the gravitational coupling
appears in the factor $GM$, and hence we cannot distinguish a
variation in $G$ from the existence of some type of dark matter.

\section{ASYMPTOTIC FREEDOM OF THE GRAVITATIONAL COUPLING}

The main idea behind the results of Refs. \cite{Goldman,Bertolami}
is the scale depedence of the gravitational coupling.
The inspiration for this comes from the property of asymptotic freedom
exhibited by 1-loop higher--derivative quantum gravity models \cite{Julve}.
Since there exists no screening mechanism
for gravity, asymptotic freedom may imply that quantum  gravitational
effects act on macroscopic and even on cosmological scales, a fact which
has of course some bearing on the dark matter problem
\cite{Goldman} and on the large scale structure of the
Universe \cite{Bertolami,Bertolami1}. It is in this framework that
a power spectrum which is consistent with the observations of IRAS
\cite{IRAS} and COBE \cite{COBE} can be obtained \cite{Bertolami,Bertolami1}.

We briefly outline this proposal.
Removing the infinities generated by quantum fluctuations and ensuring
renormalizability of a quantum field theory requires a scale--dependent
redefinition of the physical parameters.
Furthermore, the removal of those infinities still leave the physical
parameters with some dependence on finite quantities whose particular
values are arbitrary. These can be assigned by specifying
the value of the physical parameters at some momentum or length scale;
once this is performed, variations on scale are accounted for by
appropriate changes in the values of the physical parameters via
the renormalization group equations (RGEs). Thus, the
equations of motion in the quantum field theory of gravity should be
similar to the ones of the classical theory, but with their parameters
replaced by the corresponding `improved' values, that are solutions
of the corresponding RGEs. However, since gravity couples coherently to
matter and exhibits no screening mechanism, quantum fluctuations of the
gravitational degrees of freedom contribute on all scales. One must
therefore include the effect of these quantum corrections into the
gravitational coupling, $G$, promoting it into a scale--dependent
quantity. One-loop quantum gravity models indicate that the coupling
$G (\mu^2 / \mu_*^2  \sim r_*^2/r^2)$ is asymptotically free, where
$\mu_*$ is a reference momentum, meaning that $G$ grows with scale
\cite{Julve}. A typical solution for $G(r_*^2 / r^2)$ was obtained in
Ref. \cite{Goldman} setting the $\beta$-functions of matter to vanish
and integrating the remaining RGEs:
\begin{equation}\label{GR}
G(r_*^2 / r^2)=G_{lab}\, \delta(r,r_{lab})\ ,
\end{equation}
where $G_{lab}$ is the value of $G$ measured
in the laboratory at a length scale $r_{lab}$, and
$\delta(r,r_{lab})$ is a growing function of $r$. In order for the
asymptotic freedom of $G(\mu^2 / \mu_*^2)$ to have an effect
on for instance the dynamics of galaxies and their rotation curves,
the function $\delta(r,r_{lab})$ should be close to one for $r <
1$ kpc, growing significantly only for $r \geq 1$ kpc.
A convenient parametrization for $\delta(r,r_{lab})$ from
the fit of Ref. \cite{Goldman} in the kpc range is the following:
\begin{equation}\label{DEL}
\delta(r,r_{lab}) = 1.485 \left[1 +\beta \left({r \over r_0}\right)^{\gamma}
\ln({r \over r_0})\right]\ ,
\end{equation}
where $\beta \simeq 1/30$, $\gamma \simeq 1/10$ and $r_0 = 10$ kpc.

We mention that a scale dependence of the gravitational constant also arises
from completely different reasons in the so-called
stochastic inflation formalism \cite{LLM} and that the scaling behaviour
and screening of the cosmological constant was also discussed in the
context of the quantum theory of the conformal factor in four dimensions
\cite{AMM}.

In what follows we shall use the fit (1),(2) in our analysis
of the rotation curves of galaxies, and extract a prediction
for the distribution of dark matter. However, before we pursue
this discussion let us present some of the ideas developed in Refs. [5-7].
As discussed above, the classical equations
have to be `improved' by introducing the scale dependence of the
gravitational coupling. This method suggests that the presence of
cosmological dark matter could be replaced by an asymptotically free
gravitational coupling. Assuming that the Friedmann equation describing
the evolution of a flat Universe is the improved one, then:
\begin{equation}\label{H2}
H^2(\ell) = {8 \pi \over 3} G(\frac{a_0^2
 \ell_*^2}{a^2 \ell^2}) \rho_{m}\ ,
\end{equation}
where $a=a(t)$ is the scale factor, $H=\dot{a}/a$ is the Hubble
parameter, $\rho_m$ is the density of
matter, $\ell$ is the comoving distance and
$\ell_*$ is some convenient length scale.

{}From Eq. (\ref{DEL}) one sees that the
present physical density parameter, $\Omega_0^{phys}$, is by construction equal
to one. However, the quantity which is usually referred to as density
parameter is actually:
\begin{equation}\label{OME}
\Omega_0 = {8 \pi \over 3} {G \rho_{m_0} \over {H_{0_*}^2}}\ ,
\end{equation}
where $H_{0_*}$ is the present Hubble parameter for a
given large scale distance, $r = r_*$. This leads to
$\Omega_0$ as a growing function of scale, which is in
agreement with observations for a constant $\rho_{m_0}$.

Furthermore, from Eq. (3) one can clearly see the scale dependence of
the Hubble parameter \cite{Bertolami,Bertolami1,Kim}. Moreover,
as shown in Refs. \cite{Bertolami} and \cite{Bertolami1},
the power spectrum resulting from these considerations is similar to
that of a low density Cold Dark Matter model with a large cosmological
constant \cite{Efs1}.

\section{ROTATION CURVES OF GALAXIES}

Let us now turn to the discussion of the implications of the fit
(\ref{DEL}) for the rotation curves of galaxies. It is a quite well
established observational fact that the rotation curves of spiral
galaxies flatten after about 10 to 20 kpc from their centre, which of
course is a strong dynamical evidence for the presence of dark matter
and/or of non-Newtonian physics. The rotation velocity of the galaxy
is given by the non-relativistic relation,
\begin{equation}\label{V2}
v^2 =  {G(r) M(r) \over r} \ ,
\end{equation}
which approaches a constant value some distance from the centre,
{\it e.g.} $v_0^2 = 220$ km/s for the Milky Way. Assuming that the
gravitational coupling is precisely Newton's constant $G_N$ and
imposing that the rotation velocity is constant, using the Virial
Theorem at $r = R \equiv 500$ kpc, one finds the standard expression
for the mass distribution of dark matter:
\begin{equation}\label{MNR}
M_N(r) = M_N(R) {r \over R}\ .
\end{equation}
Assuming instead a running gravitational coupling (\ref{DEL}), the
condition that the rotation velocity is constant yields:
\begin{equation}\label{MR}
M(r) = {0.673 \over \left[1 + \beta ({r \over r_0})^{\gamma}
ln({r\over r_0})\right]} M_N(r)\ .
\end{equation}
Equation (\ref{MR})
reveals after simple computation that the running of the gravitational
coupling reduces the amount of dark matter required to explain the flatness
of the rotation curves of galaxies by about $44 \%$, assuming that galaxies
stretch up to a distance of about 500 kpc. This result \cite{BGB}
(see also Ref. \cite{BKS}) is a clear prediction of the dependence of the
gravitational coupling with scale and, in particular, of the fit (\ref{DEL}).
Furthermore, since the possibility that the Galactic halo is entirely made
up of baryonic dark matter is barely consistent with the nucleosynthesis
bounds on the amount of baryons \cite{CST}, the running of $G$ is quite
welcome since it reduces the required amount of baryonic dark matter in the
halo. An entertaining hypothesis could be that
precisely this effect is responsible for the reduction in the microlensing
event rates across the halo in the direction of the Large Magellanic Cloud
\cite{Alc,Aub} with respect to those along the bulge of our galaxy,
as reported by \cite{Udalski}.

\section{BOUNDS ON THE VARIATION OF $G$ WITH SCALE}

In this section we constrain the variation of the gravitational coupling
given by the fit (\ref{DEL}) with bounds from primordial nucleosynthesis,
binary pulsars and gravitational lensing and also discuss the effect that
a scale-dependent $G$ has on the peculiar velocity field \cite{BGB}.

\subsection{Primordial nucleosynthesis}

As mentioned in the introduction, one could obtain bounds on the
variation of the gravitational coupling from observations of the light
elements' abundances in the Universe. Such observations are in
agreement with the standard primordial nucleosynthesis scenario,
but there is still some room for
variations in the effective number of neutrinos, the baryon fraction
of the universe and also in the value of the gravitational constant.
For instance, the predicted mass fraction of primordial $^4${\it He}
can be parametrised, in theories with a variable gravitational coupling,
in the following way \cite{OSSW},
\begin{equation}\label{PNS}
Y_{\rm p} = 0.228 + 0.010 \ln \eta_{10} + 0.327 \log \xi\ ,
\end{equation}
where $\eta_{10}$ is the baryon to photon ratio in units of $10^{-10}$
and $\xi$ is the ratio of the Hubble parameter at nucleosynthesis and
its present value, itself proportional to the square root of the
corresponding gravitational constant. In the fit
(\ref{PNS}) it is assumed that the effective number of light neutrinos
is $N_\nu = 3$  and that the neutron lifetime is $\tau_n =
887$ seconds.

By running the nucleosynthesis codes for different values of $G$,
it was shown in Ref. \cite{AKR} that a variation of $\Delta G/G = 0.2$
on the values of the gravitational coupling was compatible with the
observations of the primordial $D$, $^3${\it He}, $^4${\it He} and
$^7${\it Li} abundances at $1\sigma$ level.

This result will now be used to constrain the running of $G$ in an
asymptotically free theory of gravity. In a theory
with a scale-dependent
gravitational constant, the maximum value of $G$ at a given time is the
one that corresponds to the physical horizon distance at that time.
During primordial nucleosynthesis, the horizon distance grows from a few
light-seconds to a few light-minutes, {\it i.e.} less than a few
milliparsecs. At that scale we find $\Delta G/G = 0.07$, see Eq.
(\ref{DEL}), which is much less than the allowed variation of $G$
given in \cite{AKR}. Therefore, primordial nucleosynthesis does not rule
out the possibility of an asymptotically free gravitational coupling.
Of course, a light-second is about the distance to the Moon, and there
are similar constraints on a variation of $G$ at this scale coming from
lunar laser ranging, $\Delta G/G < 0.6$ \cite{TEGP}.

\subsection{Binary pulsars}

The precise timing of the orbital period of binary pulsars and, in
particular, of the pulsar PSR 1913+16, provides another way of obtaining
a model-independent bound on the variation of the gravitational coupling
\cite{TW}. Since the semimajor axis of that system is just
about a few light-seconds, the resulting limits on the variation of $G$
can be readily compared with the ones arising from nucleosynthesis.
The observational limits on the rate of change of the orbital period,
mainly due to gravitational radiation damping, together
with the knowledge of the relevant Keplerian and post-Keplerian orbiting
parameters, allows one to obtain the following limit \cite{TW}:
\begin{equation}
\sigma \equiv {\Delta G\over G} < 0.08\,h^{-1} \ ,
\end{equation}
where $h$ is the value of the Hubble parameter in units of 100 km/s/Mpc.
For $h = 0.8$, \cite{Freed} one obtains $\sigma = 0.1$ which is more
stringent than the nucleosynthesis bound, but is still compatible with the fit
(\ref{DEL}).

\subsection{Gravitational lensing}

Gravitational lensing of distant quasars by intervening galaxies may provide,
under certain assumptions, yet another method of constraining, on large
scales, the variability of the gravitational coupling. The four observable
parameters associated with lensing, namely, image splittings, time delays,
relative amplification and optical depth do depend on $G$, more precisely on
the product $GM$, where $M$ is the mass of the lensing object. This dependence
might suggest that limits on the variability of $G$ could not be obtained
before an independent determination of the mass of the lensing object.
However, as the actual bending angle is not observed directly, the relevant
quantities are the distance of the lensing galaxy and of the quasar. Since
these quantities are inferred from the redshift of those objects, they depend
on their hand on $G$, on the Hubble constant, $H_0$, and on the density
parameter, $\Omega_0$. However, as we have previuosly seen, a
scale-dependent gravitational coupling implies also a dependence on scale of
$H_0$ and $\Omega_0$, see Eqs. (3) and (4). This involved dependence
on scale makes it difficult to proceed as in Ref. \cite{LENS},
where gravitational lensing in a flat, homogeneous and isotropic cosmological
model, in the context of a Brans-Dicke theory of gravity, was used to provide
a limit on the variation of $G$:
\begin{equation}\label{GLENS}
{\Delta G\over G} = 0.2 \ .
\end{equation}
Since for this limit $\Omega_0 = 1$ was assumed, while in a scale-dependent
model it is achieved via the running of the gravitational coupling, the
bound (\ref{GLENS}) contrains only residual variations of $G$ that have not
been already taken into account when considering the dependence on scale
of $H_0$ and $\Omega_0$. Of course, for models where the cosmological
parameters are independent of scale, the bound (\ref{GLENS}) can be readily
used to constrain the variability of $G$ on intermediate cosmological scales.
It is worth stressing that this method, besides being one of the few
available where this variability is directly constrained at intermediate
cosmological times between the present epoch and the nucleosynthesis era, it
is probably the only one which can realistically provide in the near future
even more stringent bounds on even larger scales by observing the lensing
of light from far away quasars caused by objects at redshifts of order
$z \geq 1$.

\subsection{Peculiar velocity field}

Since we expect the effects of a running $G$ to become important at very
large scales, one could try to explore distances of hundreds of Mpc,
where the gravitational coupling is significantly different from that of
our local scales. That is the realm of physical cosmology where of
particular importance is the study of the  peculiar
velocity field. A possible signature of the running of $G$ would be a
mismatch between the velocity fields and the actual mass distribution,
such that at large scales the same mass would pull more strongly.
To be more specific, in an expanding universe there is a relation
between the kinetic and gravitational potential energy of density
perturbations known as the Layzer-Irvine equation
(see eg.Ref.  \cite{Peebles}) that can be written as a relation
between the mass-weighted mean square velocity $\bar v^2$ and the mass
autocorrelation function $\xi(r)$,
\begin{equation}\label{LI}
\bar v^2 (r) = 2\pi G~ \rho_b~ J_2(r)\ ,
\end{equation}
where $\rho_b$ is the mean local mass density and
$J_2(r) = \int_0^r r~dr~\xi(r)$. The galaxy-galaxy correlation function
can be parametrized by $\xi(r) \sim (r/r_0)^{-1.8}$ with $r_0 = 5 h^{-1}$
Mpc, while the cluster-cluster correlation function has the same
expression with $r_0 = 20 h^{-1}$ Mpc. This means that the velocity
field (\ref{LI}) should be proportional to $(r/r_0)^{0.2}$, unless the
gravitational constant has some scale dependence. So far the relation
seems to be satisfied, under rather large observational errors (see
Ref. \cite{Dekel} for a review). Unfortunately, the
errors are so large that it would be premature to infer from this a scale
dependence of $G$. Even worse, phenomenologically there is a
proportionality constant between the galaxy-galaxy correlation
function and the
actual mass correlation function, the so- called biasing factor, which is
supposed to be scale dependent and could mimic a variable gravitational
constant. However, future sky surveys might be able to constrain more
strongly the relation (\ref{LI}) by measuring peculiar velocities with
better accuracy at larger distances and it might then be possible to extract
the scale-dependence of $G$.

\section{CONCLUSIONS}

We have seen that the running of the gravitational
coupling is compatible with the observational fact that the rotation
curves of galaxies are constant provided some amount of baryonic dark matter
is allowed, actually about $44 \%$ less than what is required for a
constant $G$. This could also explain why we see less microlensing events
towards the halo than in the direction of the bulge of our galaxy.
Failure in reproducing the predicted distribution of baryonic dark
matter would signal either that the approach adopted here is unsuitable
or that the fit (\ref{DEL}) is inadequate. We have looked for possible
bounds on variations of $G$ with scale from primordial nucleosynthesis,
variations in the period of binary pulsars,
macroscopic gravitational lensing and deviations in the peculiar
velocity flows. Unfortunately, as observational errors tend to increase
with the scale probed, we cannot yet seriuosly constrain an increase of
$G$ with scale, as proposed by the asymptotically free theories of gravity.

\end{document}